\documentclass[aps,prb,groupedaddress,twocolumn]{revtex4-1}
\usepackage{dcolumn}
\usepackage{bm}
\usepackage[colorlinks,hyperindex, urlcolor=blue, linkcolor=blue,citecolor=black, linkbordercolor={.7 .8 .8}]{hyperref}
\usepackage{graphicx}
\usepackage{tabularx}
\usepackage{amsfonts}
\usepackage{amsmath}
\usepackage{amssymb}
\usepackage{amsbsy}
\usepackage{setspace}
\usepackage{nicefrac}
\usepackage{xfrac}
\usepackage{bm}
\usepackage{float}

\begin{document}

\title{Unexpected hole doping of graphene by osmium adatoms}

\author{Jamie A.~Elias}
\affiliation{Department of Physics, Washington University in St.~Louis, 1 Brookings Dr., St.~Louis MO 63130, USA}
\author{Erik~A.~Henriksen}
\email{henriksen@wustl.edu}
\affiliation{Department of Physics, Washington University in St.~Louis, 1 Brookings Dr., St.~Louis MO 63130, USA}
\affiliation{Institute for Materials Science \& Engineering, Washington University in St.~Louis, 1 Brookings Dr., St.~Louis MO 63130, USA}

\begin{abstract} The electronic transport of monolayer graphene devices is studied before and after \emph{in situ} deposition of a sub-monolayer coating of osmium adatoms. Unexpectedly, and unlike all other metallic adatoms studied to date, osmium adatoms shift the charge neutrality point to more positive gate voltages. This indicates that osmium adatoms act as electron acceptors and thus leave the graphene hole-doped. Analysis of transport data suggest that Os adatoms behave as charged impurity scatterers, albeit with a surprisingly low charge-doping efficiency. The charge neutrality point of graphene is found to vary non-monotonically with gate voltage as the sample is warmed to room temperature, suggesting that osmium diffuses on the surface but is not completely removed.
\end{abstract}
%\shortabstract

%\begin{document}
\maketitle

\section{Introduction}

Graphene, an atomically-thin honeycomb lattice of carbon atoms, hosts an electronic system that is inherently unprotected from external influences~\cite{castro_neto_electronic_2009}. Thus, the electronic behavior can be readily altered by proximity to supporting substrates and incidental adsorbates~\cite{chen_charged-impurity_2008,jang_tuning_2008}. While such effects generically result in a more disordered or lower mobility electronic system, there is nonetheless much interest in the potential to use surface adsorbates to advantageously alter the electronic properties of graphene. For instance adatoms may yield unusual surface ordering~\cite{cheianov_hidden_2009,shytov_long-range_2009}, atomic collapse states~\cite{shytov_atomic_2007,wang_observing_2013}, and gate-voltage control of magnetism~\cite{wehling_orbitally_2010,pike_graphene_2014}; for a review of approaches to develop novel insulating or magnetic states in graphene see Ref.~\cite{katoch_adatom-induced_2015}.

A particularly intriguing goal for adatom-decorated graphene is to induce a greater spin-orbit coupling than the native and usually negligible coupling due to C atoms~\cite{min_intrinsic_2006,yao_spin-orbit_2007,gmitra_band-structure_2009,sichau_resonance_2019}. Indeed, both the quantum spin~\cite{kane_quantum_2005} and quantum anomalous~\cite{haldane_model_1988,liu_quantum_2008} Hall effects are predicted to arise in graphene decorated with heavy elements having a strong intrinsic spin-orbit coupling~\cite{qiao_quantum_2010,tse_quantum_2011,weeks_engineering_2011,hu_giant_2012,zhang_electrically_2012}. As always the devil is in the details: for instance the indium and thallium atoms discussed in Ref.~\cite{weeks_engineering_2011} are expected (and for indium, observed~\cite{chandni_transport_2015,jia_transport_2015}) to donate electrons to graphene. Thus for the required adatom density of a few percent of a monolayer, the Fermi level will have moved far from the resulting topological gap. This leaves the gap inaccessible to measurement by electronic transport, even with techniques for inducing large electric field effects~\cite{efetov_controlling_2010}. In contrast, Ref.~\cite{hu_giant_2012} predicts that osmium adatoms will induce a giant topological gap of order 0.2 eV, and the Fermi level will naturally reside in this gap. This has the exciting potential to yield a robust quantum spin Hall system, and is the primary motivation for this work.

Here we report electronic transport experiments on three monolayer graphene devices onto which we evaporate a dilute coating of Os adatoms, deposited \emph{in situ} in the ultra-high-vacuum environment of a cryostat. We observe a customary increase in charge carrier scattering, but unexpectedly find the charge neutrality point in graphene to shift to more positive gate voltages, indicating the graphene is becoming hole-doped. This runs counter to expectations~\cite{nakada_dft_2011,hu_giant_2012,manade_transition_2015}, and to all other metal atoms placed on graphene to date including K, Pt, Fe, Ti, Gd, Ca, In, Mg, Au, Ir, and Li~\cite{chen_charged-impurity_2008,pi_electronic_2009,alemani_effect_2012,katoch_impact_2012,chandni_transport_2015,jia_transport_2015,wang_neutral-current_2015,wang_electronic_2015,khademi_alkali_2016}. 

\begin{figure}[t]
\includegraphics[width=\columnwidth]{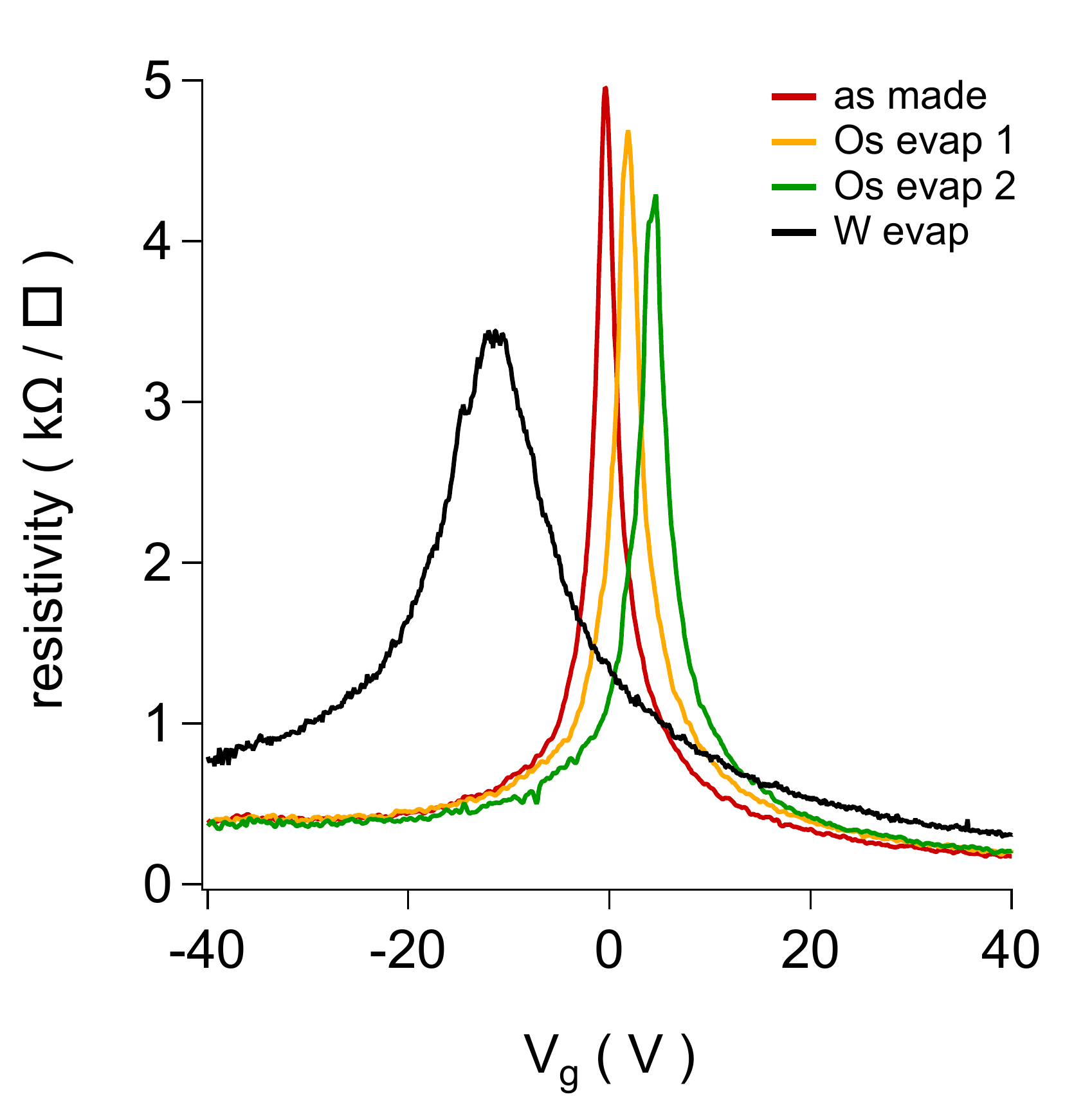}
\caption{The resistivity peak in a monolayer graphene device (D1) shifts to the right after depositing Os adatoms, and to the left for W adatoms~\cite{elias_electronic_2017}, indicating $p$-type and $n$-type doping, respectively. Evaporations were done sequentially in a single experimental run, by exhausting all Os from a single evaporation source and then increasing the current until the W source wire itself began to evaporate. \label{ostow} } 
\end{figure}

Figure~\ref{ostow} illustrates the central result of this work. A monolayer graphene device, D1, was exposed to an Os-covered W wire used as the evaporation source. By passing a current through the wire, Os atoms are caused to evaporate with some landing on the graphene. In two successive evaporations, the charge neutrality point indicated by the resistance maximum (``Dirac peak'') is seen to shift upward in gate voltage, indicating that the graphene has become hole-doped. After exhausting the Os source, the current is increased until W atoms from the source wire are evaporated. These donate electrons to graphene, as we have previously observed~\cite{elias_electronic_2017}. Thus we can clearly distinguish the charge-doping behavior of Os from W adatoms.

\section{Experiment}

\begin{figure}[!b]
\includegraphics[width=\columnwidth]{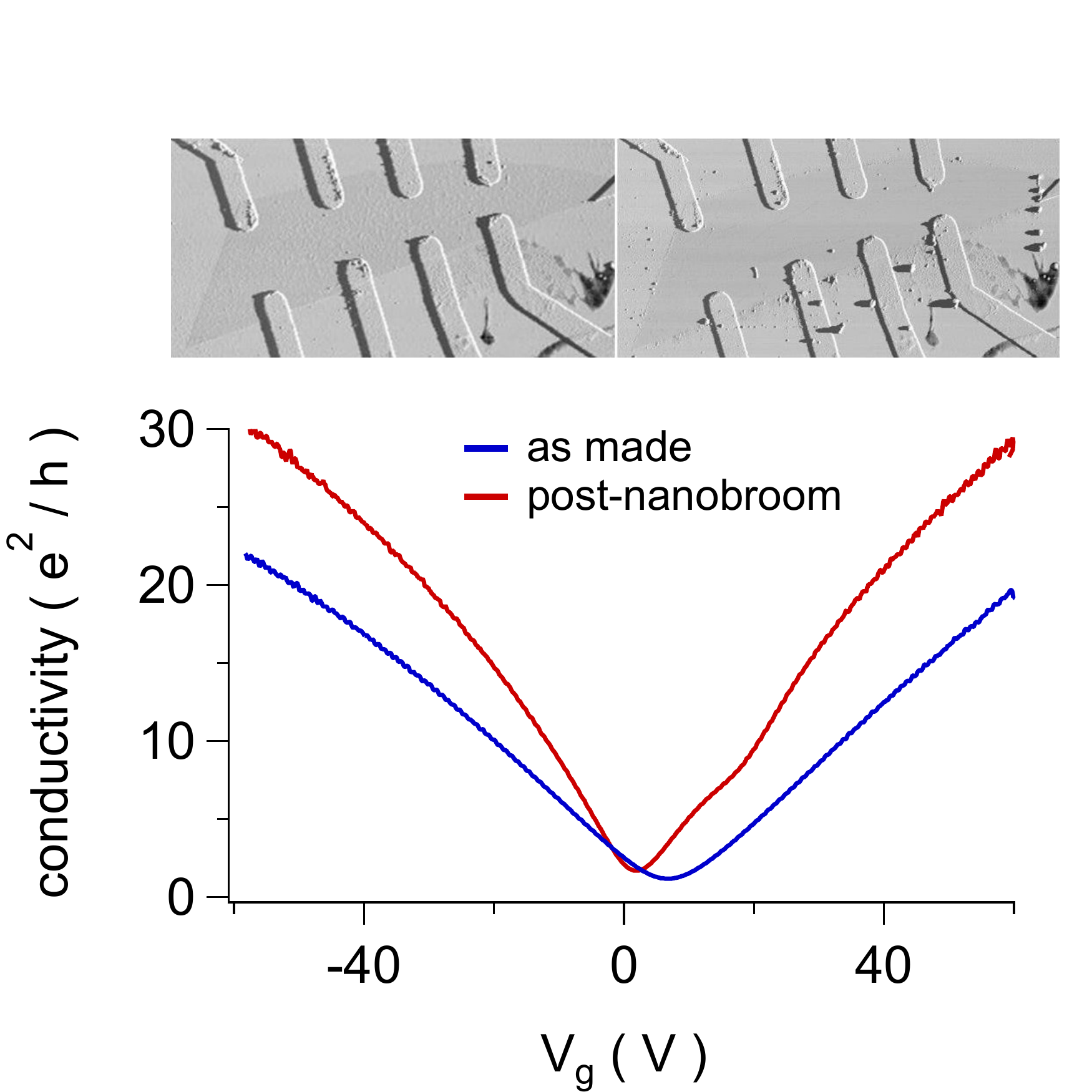}
\caption{Top: monolayer graphene-on-SiO$_2$ device before (left) and after (right) nano-brooming via contact atomic force microscopy. A significant decrease in roughness is observed after sweeping remnant fabrication residues away~\cite{goossens_mechanical_2012,lindvall_cleaning_2012}. Bottom: the brooming procedure results in improved transport, as the mobility is linearly dependent on the conductivity. \label{broom}} 
\end{figure}

We work with mechanically exfoliated monolayer graphene devices supported on SiO$_2$, with several Cr/Au metal contacts defined by standard electron beam lithography and thin film evaporation. The wafers are pre-cleaned using an O$_2$ plasma and exfoliation is performed on a hotplate to increase the size and yield~\cite{huang_reliable_2015}. For most devices, after fabrication the remnant polymer residues~\cite{lin_graphene_2012} were removed by contact atomic force microscopy ``nano-brooming'', rastering a tip at 40~nN force and 2 Hz cycle to physically sweep the surface clean~\cite{goossens_mechanical_2012,lindvall_cleaning_2012}. Figure~\ref{broom} shows AFM images of a typical device before and after brooming, along with a concomitant improvement in the quality of electronic transport. Electronic transport was also measured in devices with contacts directly defined by a shadow mask so that no residues needed to be removed; results in these devices are consistent with those that were broomed. Here we focus on results from three devices: D1, in Fig~\ref{ostow}, was used to verify that the prior W adatom electron-doping results were reproducible after the observation of hole-doping by Os adatoms. Device D2 is a monolayer graphene-on-SiO$_2$ with lithographically-defined contacts, and was broomed after fabrication. Device D3 is a monolayer graphene-on-SiO$_2$ with shadow-mask-defined contacts.

\begin{figure*}[!tbp]
\begin{center}
\includegraphics[width=0.9\textwidth]{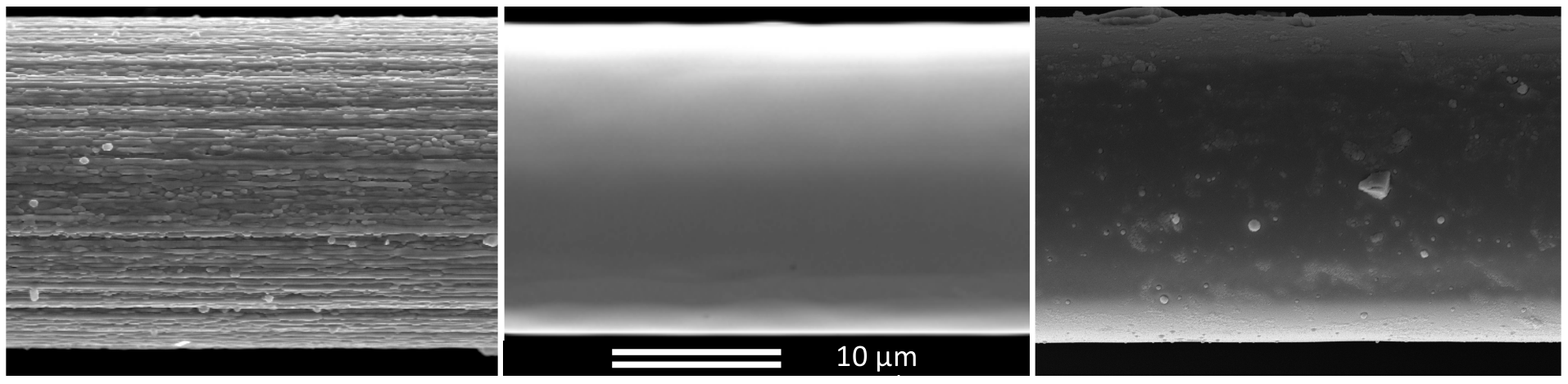}
\caption{Left: as-supplied W wire with nominal $20~\mu$m diameter (actual diameter here $19.6~\mu$m) for use as evaporation source. Middle: same wire, after thermal annealing in vacuum by passing 200 mA for 3 minutes~\cite{langmuir_vapor_1913}, which reduced the diameter to $18.4~\mu$m in this case. Right: source wire after coating with Os metal by plasma vapor deposition to a final diameter of $19.1~\mu$m. \label{sources}} 
\end{center}
\end{figure*}

Electronic transport measurements were carried out in a cryostat equipped with a 14 T superconducting solenoid, with samples mounted on a custom-built stage with independent temperature control. The stage faces downward toward a small thermal evaporator with source support posts machined from tantalum rods, which has a very high vapor pressure so that even at the high temperatures required to evaporate other refractory metals, the supports will not contaminate the deposition. We employ fine W wire sources for evaporation so that high temperatures can be reached while the overall heat load is minimized. As shown in Fig.~\ref{sources}, the as-supplied $20~\mu$m W wires are annealed by passing a current through the wire in vacuum, and later coated with the desired evaporant; in this case, osmium metal deposited from a plasma source~\cite{noauthor_courtesy_nodate}. During evaporation the sample stage temperature slowly increases from 4 K to  between 20 and 40 K depending on how long the evaporation is carried out. After deposition the sample is allowed to cool back to 4 K before proceeding with standard low-frequency lockin measurements of electronic transport. In the primary device studied here, D2, we alternate evaporation steps with annealing cycles, where the sample temperature is cycled to 270 K and back to 4 K for measurement.

\section{Results}

\begin{figure}[!bp]
\includegraphics[width=\columnwidth]{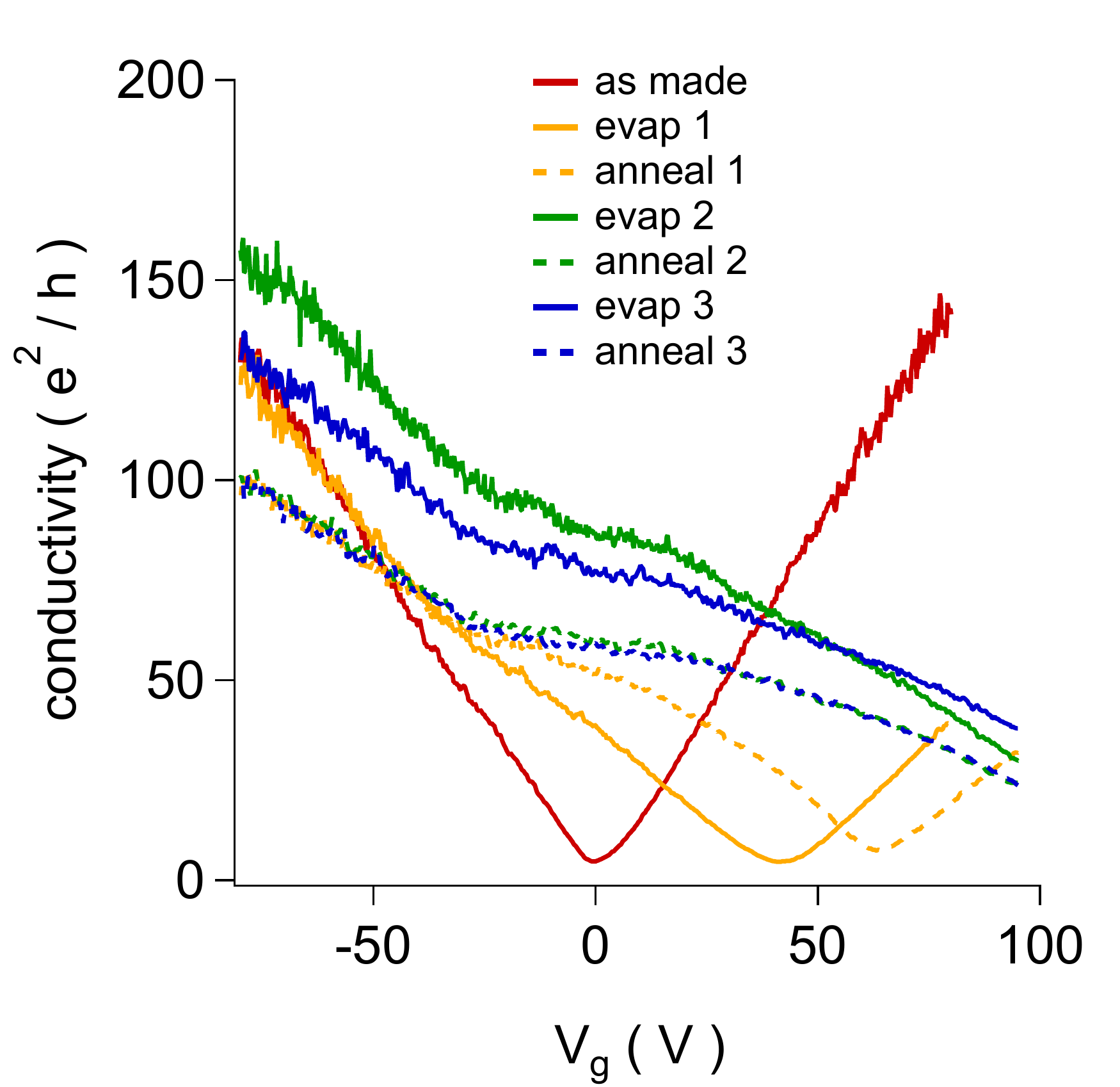}
\caption{Conductivity vs V$_g$ in a second monolayer graphene device (D2) with successive Os depositions and thermal annealing cycles. The four-terminal transport is measured immediately after each deposition and annealing cycle; the latter takes the sample to 270 K and back to 4 K prior to measurement.  \label{cond}} 
\end{figure}

In Figure~\ref{cond} we show the conductivity vs back gate voltage in a monolayer graphene device for a series of evaporations and thermal anneal cycles. The as-made device shows essentially no extrinsic doping, with the conductivity minimum $\sigma_{min}$ located at V$_g{=}0$. With each successive evaporation, the minimum conductivity is seen to shift toward positive gate voltages, indicating that the Os adatoms act as acceptors, leaving holes behind in the graphene. By the second evaporation step, the minimum conductivity has moved beyond the safe gate voltage range and can only be inferred by a linear fit to the $\sigma(\textrm{V}_g)$ relation; clearly this is somewhat complicated by the presence of a small dip in the conductivity in the neighborhood of V$_g{=}-20$ V. In the present work we simply draw a line along the conductivity data that passes over this dip, and determine the charge neutrality point location as the intersection of this line with the V$_g$ axis at $\sigma{=}0$. 

\begin{figure}[!tbp]
\includegraphics[width=\columnwidth]{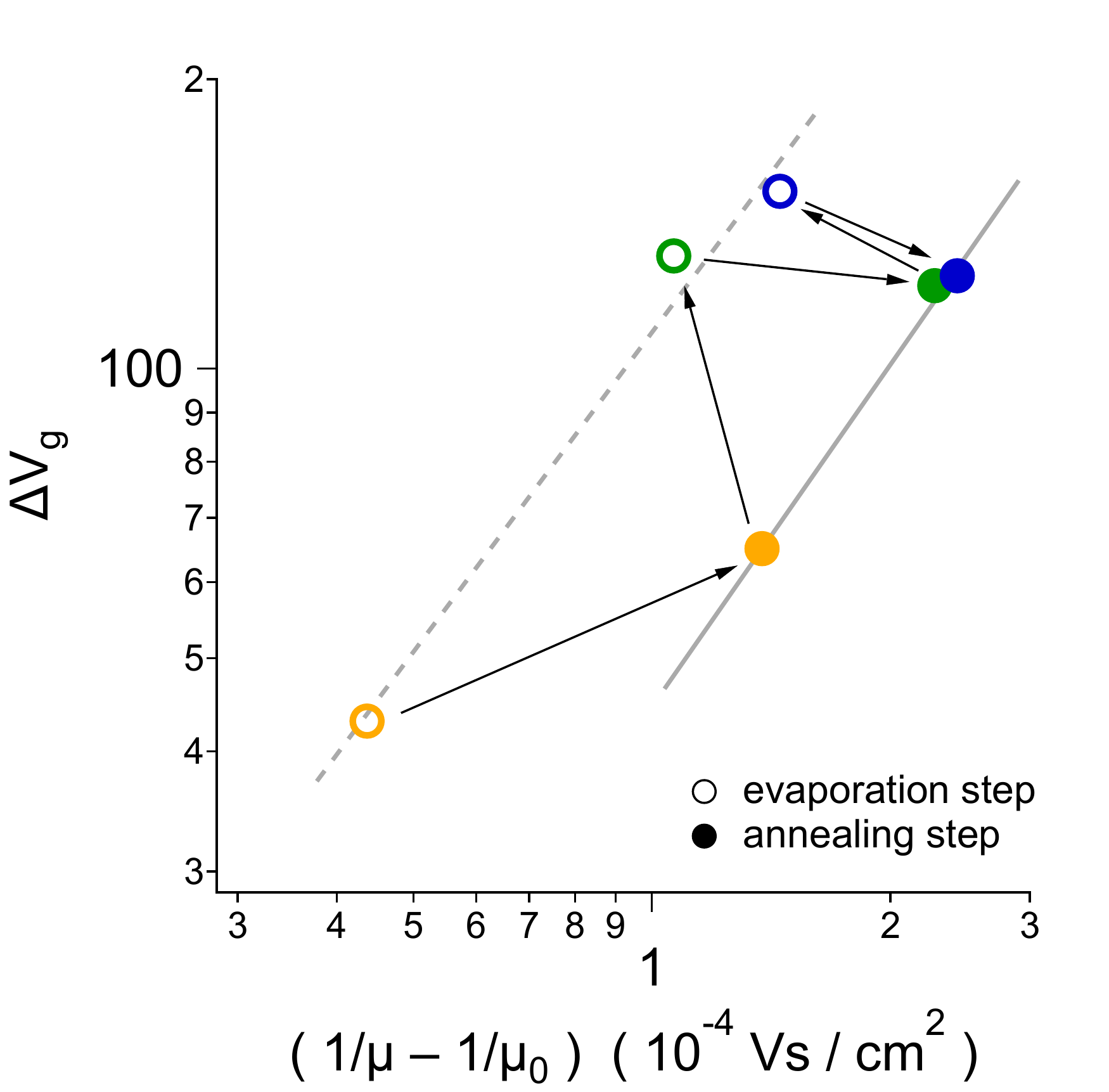}
\caption{The shift of the minimum conductivity in gate voltage vs changes in the scattering rate, as parametrized by the inverse mobility. Both are relative to the as-made device, and measured following each evaporation and annealing step in the order indicated by the arrows. The color scheme follows that in Fig.~\ref{cond}. Theory suggests a relation $\Delta$V$_g{\sim}(\mu^{-1}{-}\mu_0^{-1})^b$, with $b{\approx}1$ for Coulomb scattering~\cite{adam_self-consistent_2007,chen_charged-impurity_2008}. The solid and dashed grey lines are fits to the annealing or evaporation data only, and yield $b{=}1.2$ and $1.1$ respectively. \label{dvg}} 
\end{figure}

There are two additional features: first, the annealing steps significantly impact both the position of $\sigma_{min}$ and the 
the slope of $\sigma(\textrm{V}_g)$; and second, the slope generally becomes shallower as the experiment progresses, indicating that the device mobility is on the whole decreasing. The detailed behavior is intriguing. The first anneal shifts $\sigma_{min}$ further to the right and decreases the slope, while the following evaporation (evap 2) significantly shifts the location of $\sigma_{min}$ but barely changes the slope. The second anneal impacts the mobility---reducing the slope of $\sigma(\textrm{V}_g)$---but barely changes the location of $\sigma_{min}$. We collect these observations in Fig.~\ref{dvg} where we plot the shift of $\sigma_{min}$ (that is, the effect of charge doping) relative to the as-made device by $\Delta \textrm{V}_g{=}\textrm{V}_g(\sigma_{min}){-}\textrm{V}_{g0}$, as a function of the change in scattering rate given by the difference in the inverse mobility. Each data point is marked as either an evaporation or annealing step. A  pattern emerges in which at each step the device state moves upward to the right in a zig-zag fashion, as each step appears to preferentially impact either the charge doping or the mobility. Indeed after the initial decrease in mobility from the as-made device, the mobility decreases at each annealing step but, counterintuitively, is seen to \emph{increase} at each new evaporation.

A power law relation is often assumed between $\Delta \textrm{V}_g$ and the inverse mobility, $\Delta \textrm{V}_g{=}(\mu^{-1}{-}\mu_0^{-1})^b$, where $\mu^{-1}$ is proportional to the scattering potential. The exponent $b$ can be related to the dominant scattering mechanism, e.g.~$b\approx1$ represents Coulomb scattering by impurities located a small distance above the graphene plane~\cite{adam_self-consistent_2007,chen_charged-impurity_2008}. In Fig.~\ref{dvg} we show two fits, with the dashed (solid) line fitting just the evaporation (annealing) step data, for which $b{=}1.1$ ($1.2$). This is close to the $1.2{-}1.3$ seen for K, or $1.4$ for In, and a bit above the range of $0.64{=}1.1$ for Ti, Fe, and Pt adatoms~\cite{chen_charged-impurity_2008,pi_electronic_2009,elias_electronic_2017}. It suggests that although Os has different sign of charge doping than other metals, the mechanism is similar in that the ionized Os adatoms act as Coulomb scatterers.

\begin{figure}[!b]
\includegraphics[width=\columnwidth]{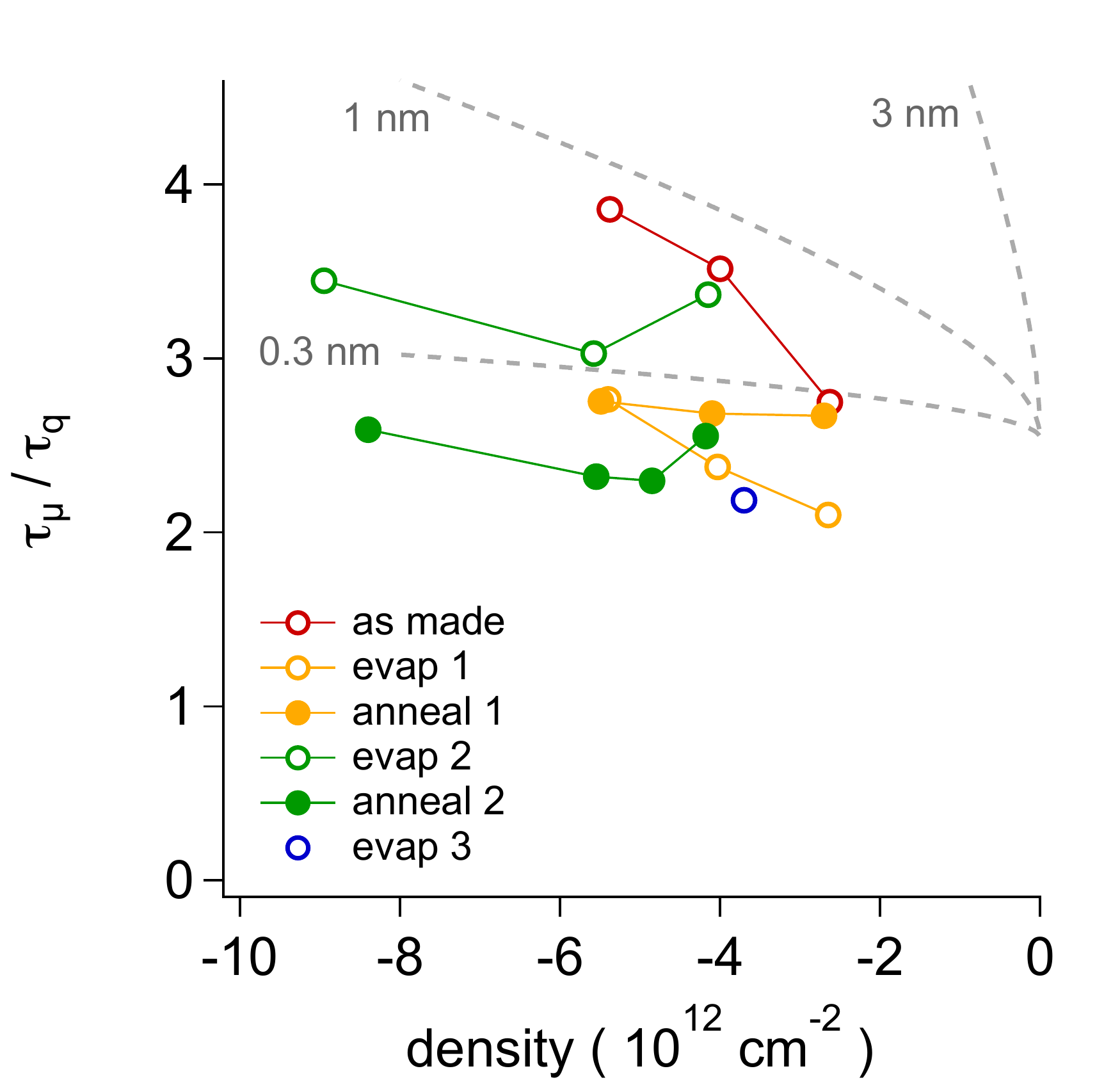}
\caption{The ratio of the transport and quantum scattering times vs carrier density, for evaporation and anneal steps. An overall decrease of roughly 30\% is seen. The dashed lines are theoretical predictions for Coulomb impurities located a distance $d{=}0.3$, 1, and 3 nm above the plane~\cite{hwang_single-particle_2008}. \label{ratio}} 
\end{figure}

Next we examine the scattering times extracted from the electronic transport. From the conductivity we extract the momentum scattering time $\tau_\mu{=}\sigma/(2 e^2 k_F v_F/h)$, and from the envelope of the Shubnikov-de Haas oscillations observed in finite magnetic field, we extract the quantum scattering time, $\tau_q$. The details of this procedure are discussed in Ref.~\cite{elias_electronic_2017}. The ratio $\tau_\mu/\tau_q$ of these characteristic times, commonly used to distinguish between scattering mechanisms~\cite{das_sarma_single-particle_1985,hwang_single-particle_2008}, is shown in Fig.~\ref{ratio} as a function of carrier density for the evaporation and annealing steps. The ratio shows the same sort of back-and-forth behavior apparent in Fig.~\ref{dvg} due to the mobility decreasing for anneals yet increasing after evaporations. On the whole, the ratio decreases by roughly a third from its initial values in the as-made device. We use a theory of Hwang \& das Sarma to predict the ratio for scattering by isolated Coulomb impurities located a distance $d$ above a perfectly flat graphene plane, for $d=0.3, 1$ and 3 nm~\cite{hwang_single-particle_2008}. Interestingly, the data suggest that Os adatoms lie very close to the graphene surface, as expected from density functional theory calculations~\cite{nakada_dft_2011}.

\begin{figure}[tbp]
\includegraphics[width=\columnwidth]{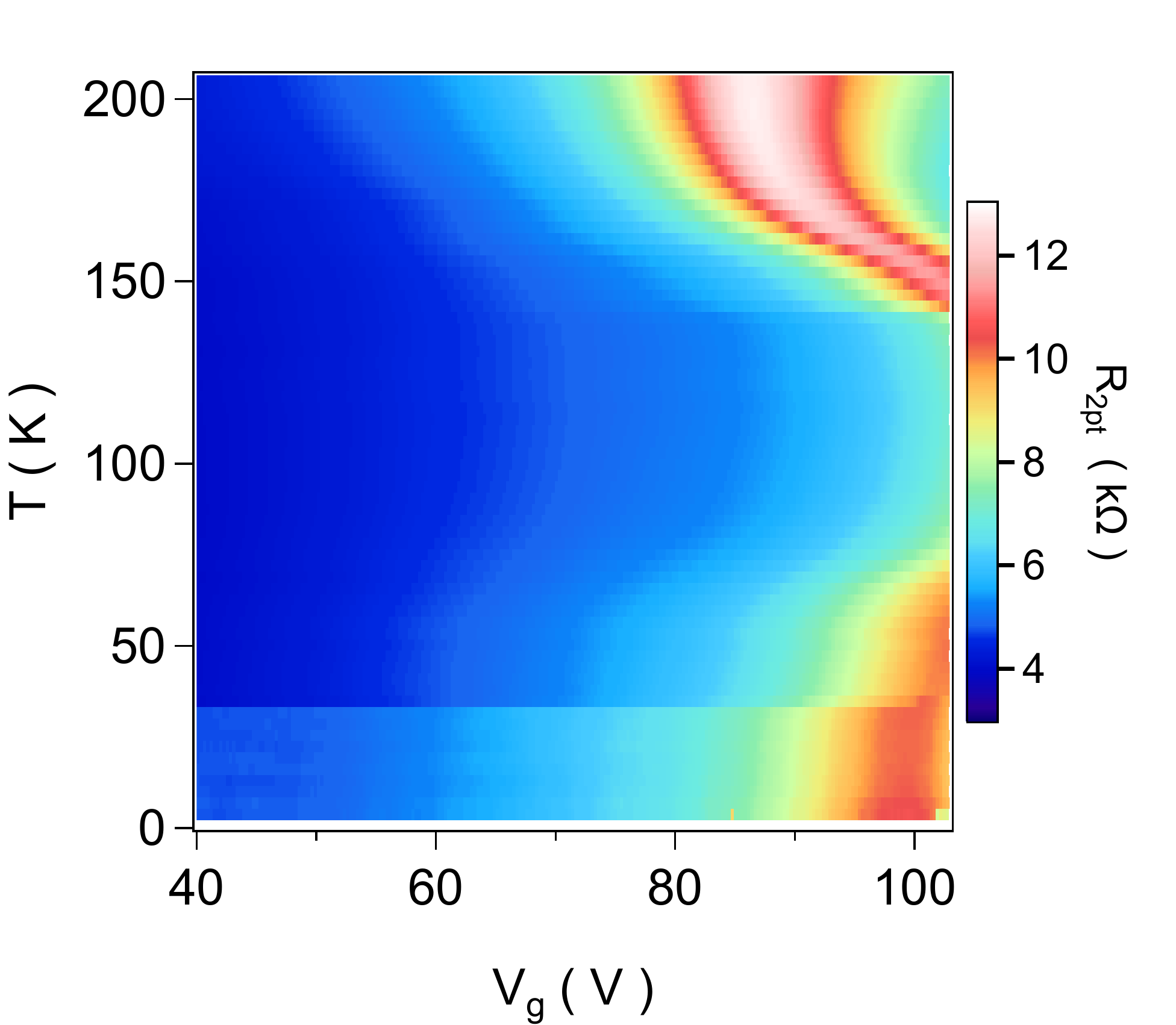}
\caption{Resistance vs V$_g$ during warmup from 2 to 200 K in a third device after Os deposition. The minimum conductivity point shifts non-monotonically with temperature outside the available gate voltage range above 60 K, and surprisingly is seen to return in view above 140 K before stabilizing with further warming. \label{warm}} 
\end{figure}

Finally we examine the annealing behavior. In prior studies of K and In adatoms, it has been found that the effect of adatom deposition is reversed upon warming to room temperature, although whether by desorption or migration to the edge or into clusters is not known~\cite{yan_correlated_2011,chandni_transport_2015}. Osmium adatoms show a distinct behavior, illustrated in Fig.~\ref{warm} in a plot of the 2-terminal resistance of a third device, D3: as the sample temperature is raised, the location of $\sigma_{min}$ shifts first to \emph{higher} gate voltages---and in fact moves beyond the accessible V$_g$ range---before returning into view above approximately 140 K. Unexpectedly, above 170 K, $\sigma_{min}$ ceases to shift and, in particular, has not returned the sample to its pristine state. During this motion, the peak resistance increases from $10.3$ k$\Omega$ to $12.7$ k$\Omega$. (Device D2 showed similar behavior but the V$_g$ range was more restricted.) Overall this behavior suggests the relation between annealing and evaporation steps seen earlier depends upon the annealing temperatures reached: annealing to lower temps will result in a significant shift in $\sigma_{min}$ to positive V$_g$, whereas annealing to higher temperatures can reverse the motion to more negative V$_g$. The reason for this motion is not yet clear. However, it surely depends in part on the site-dependent binding energies, which have been calculated by density functional theory~\cite{nakada_dft_2011,hu_giant_2012,manade_transition_2015}. While K and In are both most strongly bound at the hollow site of the graphene honeycomb, they have migration energies of 120 and 20 meV, respectively, and so at room temperature are relatively free to move. Osmium is also strongly bound to the hollow site, but with a larger barrier to migration of 750 meV. However, during evaporation the Os adatoms arrive with a random distribution over the surface, and may not necessarily fall into the hollow site. Other locations have lower binding energies, so that warming will drive the adatoms to the lower energy hollow site. This may account for the initial upshift of $\sigma_{min}$ with increasing temperature. On further increase, the Os adatoms will become mobile with a diffusion rate given by
\begin{equation*}
D=D_0~ \textrm{exp}\left(-E_m/k_B T \right)~,
\end{equation*}
where $D_0$ is a prefactor we estimate to be $7{\times}10^{-3}$ cm$^2$/s ~\cite{barth_transport_2000}, and $E_m$ is the energy barrier to migration. From this we find that Os adatoms achieve a hopping rate of one site per second right at room temperature. These estimates are very sensitive to the energy scales; if we use the data in Fig.~\ref{warm} which imply Os adatoms become mobile near 140 K, then the migration energy would be approximately 360 meV. In either case, the atoms are immobile at 50 K and below. Of course, the adatoms may also directly desorb into the vacuum, or upon migrating may rapidly join into clusters, which are likely to impact the observed behavior.

\section{Discussion}
\subsection{Os adatom density}
Each of the three evaporations in Fig.~\ref{cond} exhausted the Os supply from a W wire source with a coating that ranged from 1 to $2~\mu$m thick, as measured in a scanning electron microscope. The source wires are 1 cm long and located ${\approx}4.4$ cm away from the graphene device. A simple geometric estimate for the deposited Os adatom density can be made, yielding approximately $3{\times}10^{14}$ cm$^{-2}$ per evaporation, or an 18\% coverage (adatoms/unit cell). Meanwhile, the shift of $\sigma_{min}$ vs V$_g$ can be converted into a charge doping density via the known voltage gating efficiency for these samples, $n{=}7.2{\times}10^{10}$ cm$^{-2}$~\cite{elias_electronic_2017}, so for instance the 43 V shift of the first evaporation corresponds to an addition of $3.1{\times}10^{12}$ cm$^{-2}$ holes, or a doping efficiency of $0.01$ holes/Os adatom. This is rather lower---and of opposing sign---than the $0.3-0.6$ electrons/atom expected from DFT calculations~\cite{nakada_dft_2011,hu_giant_2012,manade_transition_2015}. 

One possibility for the low doping efficiency is that the Os adatoms form clusters upon landing on the graphene. The atoms are evaporated from a hot W wire with a temperature approaching 3000 K~\cite{langmuir_vapor_1913,honig_vapor_1969}, and likely cool by emission of phonons in the graphene, during which the adatom is able to traverse the surface. Clusters of metal atoms on graphene are known to have a much lower charge-doping efficiency~\cite{katsnelson_scattering_2009,mccreary_effect_2010}; and clustering may also inhibit the ability to observe the effects of an induced spin-orbit coupling~\cite{cresti_multiple_2014}. However, clusters will tend to favor small-angle scattering processes~\cite{katsnelson_scattering_2009} and thus one should expect an enhanced ratio of $\tau_{\mu}/\tau_q$ rather than the reduced ratio appearing in Fig.~\ref{ratio}. 

We must note a potential caveat: the Os coating on the $20~\mu$m W wires is fragile and readily flakes off, so we cannot rule out that part of the Os (though clearly not all) may have fallen off during handling and loading of the wires. In that case, the deposited density would decrease, and the doping efficiency increase.

\subsection{Hole-doping and deposition of osmium}
The ability of adsorbed molecules to cause $p$-type doping in graphene has been known for some time~\cite{schedin_detection_2007}. As noted above, however, metallic elements have only been seen to donate electrons, and DFT calculations find that Os should gain a positive charge on graphene, leaving the graphene more negatively charged. Thus an obvious concern is whether we have actually deposited osmium metal as opposed to, say, an oxide. On the whole, this seems unlikely. The chief oxide of osmium is tetravalent OsO$_4$, a highly volatile solid with a melting point of 40$^{\circ}$ C. In fact the osmium metal is plated on the W sources using plasma coating from an OsO$_4$ precursor. However if any remains on our sources, it will be rapidly pumped away during evacuation of the cryostat prior to cooling, or during the inspection by scanning electron microscope which is carried out on all sources (e.g.~to measure the initial wire diameter). The presence of Os metal on the sources was confirmed by EDX spectroscopy. Finally, the W wire sources could be a source of contamination. However, W adatoms are known to donate electrons to graphene (see Fig.~\ref{ostow}), the W is buried under the Os coating, and moreover at high temperatures the vapor pressure of W is ${\approx}20$ times lower than Os; so overall this is not a likely culprit. The vapor pressure of the Ta supports for the evaporation sources is midway between Os and W, and moreover these are far more massive (having a diameter of 0.125'' compared to $20~\mu$m) and directly coupled to the cryostat, so incidental Ta deposition is extremely unlikely. We maintain the sample stage and evaporator posts to keep them in clean condition, and have not encountered trouble with unwanted deposition in previous experiments~\cite{elias_electronic_2017}. We conclude that we are depositing osmium metal, and that osmium donates holes to graphene.

\subsection{Interpretation of transport characteristics}
Several features of the transport are consistent toward Os adatoms acting as charged impurity scattering centers: the linear relation of $\sigma$ and V$_g$, the linear relation of $\Delta \textrm{V}_g$ and $\mu^{-1}$, and the behavior of $\tau_{\mu}/\tau_q$. However, the extremely low charge doping efficiency, and the unexpected hole doping, strongly suggest that Os behaves qualitatively different from other metallic adatoms; in which case it is possible that alternatives to the basic model of scattering by charged surface dopants can explain our observations. Understanding the origin of the hole-doping will likely be key to solving this mystery.

\section{Conclusion}

In conclusion, we have performed electronic transport studies of graphene with a dilute coating of osmium adatoms. The osmium is found to donate holes to graphene, and to have a surprisingly small charge doping efficiency, both counter to expectations. We find that otherwise the graphene transport is impacted in similar fashion to other metallic adatoms, in that the scattering appears consistent with isolated Coulomb impurities a small distance above the surface. While motivated to study Os adatoms due to the potential for inducing a strong spin-orbit coupling, further experiments are planned to explore this possibility.

\begin{acknowledgements}
We gratefully acknowledge Jeff Hall at Structure Probe, Inc.~for assistance with the Os coatings. We have also benefitted from discussions with Jason Alicea, Marcel Franz, and Min-Feng Tu. We thank Boyi Zhou for assistance in sample fabrication. This work was partially supported by a Sigma Xi Grant-in-Aid of Research no.~G2017100194628139, with additional support provided by Sandia National Laboratory under contract no.~1882025. We further acknowledge support from the Institute of Materials Science and Engineering at Washington University in St.~Louis.
\end{acknowledgements}

\bibliographystyle{annalen}
%\bibliography{Os_on_graphene.bib}

\begin{thebibliography}{10}
\expandafter\ifx\csname urlstyle\endcsname\relax
  \providecommand{\doi}[1]{doi:\discretionary{}{}{}#1}\else
  \providecommand{\doi}{doi:\discretionary{}{}{}\begingroup
  \urlstyle{rm}\Url}\fi

\bibitem{castro_neto_electronic_2009}
A.~H. Castro~Neto, F.~Guinea, N.~M.~R. Peres, K.~S. Novoselov, A.~Geim,
  \emph{Reviews of Modern Physics}  \textbf{2009}, \emph{81}, 109.

\bibitem{chen_charged-impurity_2008}
J.-H. Chen, C.~Jang, S.~Adam, M.~Fuhrer, E.~D. Williams, M.~Ishigami,
  \emph{Nature Physics}  \textbf{2008}, \emph{4}, 377.

\bibitem{jang_tuning_2008}
C.~Jang, S.~Adam, J.-H. Chen, E.~D. Williams, S.~Das~Sarma, M.~S. Fuhrer,
  \emph{Physical Review Letters}  \textbf{2008}, \emph{101}, 146805,
  \doi{10.1103/PhysRevLett.101.146805}.

\bibitem{cheianov_hidden_2009}
V.~V. Cheianov, V.~I. Fal'ko, O.~Sylju{\aa}sen, B.~L. Altshuler, \emph{Solid
  State Communications}  \textbf{2009}, \emph{149}, 1499,
  \doi{10.1016/j.ssc.2009.07.008}.

\bibitem{shytov_long-range_2009}
A.~V. Shytov, D.~A. Abanin, L.~S. Levitov, \emph{Physical Review Letters}
  \textbf{2009}, \emph{103}, 016806, \doi{10.1103/PhysRevLett.103.016806}.

\bibitem{shytov_atomic_2007}
A.~V. Shytov, M.~I. Katsnelson, L.~S. Levitov, \emph{Physical Review Letters}
  \textbf{2007}, \emph{99}, 246802, \doi{10.1103/PhysRevLett.99.246802}.

\bibitem{wang_observing_2013}
Y.~Wang, D.~Wong, A.~V. Shytov, V.~W. Brar, S.~Choi, Q.~Wu, H.~Z. Tsai,
  W.~Regan, A.~Zettl, R.~K. Kawakami, S.~G. Louie, L.~S. Levitov, M.~F.
  Crommie, \emph{Science}  \textbf{2013}, \emph{340}, 734,
  \doi{10.1126/science.1234320}.

\bibitem{wehling_orbitally_2010}
T.~O. Wehling, A.~V. Balatsky, M.~I. Katsnelson, A.~I. Lichtenstein, A.~Rosch,
  \emph{Physical Review B}  \textbf{2010}, \emph{81}, 115427,
  \doi{10.1103/PhysRevB.81.115427}.

\bibitem{pike_graphene_2014}
N.~A. Pike, D.~Stroud, \emph{Applied Physics Letters}  \textbf{2014},
  \emph{105}, 052404, \doi{10.1063/1.4892573}.

\bibitem{katoch_adatom-induced_2015}
J.~Katoch, \emph{Synthetic Metals}  \textbf{2015}, \emph{210}, 68,
  \doi{10.1016/j.synthmet.2015.07.017}.

\bibitem{min_intrinsic_2006}
H.~Min, J.~E. Hill, N.~A. Sinitsyn, B.~R. Sahu, L.~Kleinman, A.~H. MacDonald,
  \emph{Physical Review B}  \textbf{2006}, \emph{74}, 165310,
  \doi{10.1103/PhysRevB.74.165310}.

\bibitem{yao_spin-orbit_2007}
Y.~Yao, F.~Ye, X.-L. Qi, S.-C. Zhang, Z.~Fang, \emph{Physical Review B}
  \textbf{2007}, \emph{75}, 041401, \doi{10.1103/PhysRevB.75.041401}.

\bibitem{gmitra_band-structure_2009}
M.~Gmitra, S.~Konschuh, C.~Ertler, C.~{Ambrosch-Draxl}, J.~Fabian,
  \emph{Physical Review B}  \textbf{2009}, \emph{80}, 235431,
  \doi{10.1103/PhysRevB.80.235431}.

\bibitem{sichau_resonance_2019}
J.~Sichau, M.~Prada, T.~Anlauf, T.~J. Lyon, B.~Bosnjak, L.~Tiemann, R.~H.
  Blick, \emph{Physical Review Letters}  \textbf{2019}, \emph{122}, 046403,
  ISSN 0031-9007, 1079-7114, \doi{10.1103/PhysRevLett.122.046403}.

\bibitem{kane_quantum_2005}
C.~L. Kane, E.~J. Mele, \emph{Physical Review Letters}  \textbf{2005},
  \emph{95}, 226801, \doi{10.1103/PhysRevLett.95.226801}.

\bibitem{haldane_model_1988}
F.~Haldane, \emph{Physical Review Letters}  \textbf{1988}, \emph{61}, 2015.

\bibitem{liu_quantum_2008}
C.-X. Liu, X.-L. Qi, X.~Dai, Z.~Fang, S.-C. Zhang, \emph{Physical Review
  Letters}  \textbf{2008}, \emph{101}, 146802,
  \doi{10.1103/PhysRevLett.101.146802}.

\bibitem{qiao_quantum_2010}
Z.~Qiao, S.~A. Yang, W.~Feng, W.-K. Tse, J.~Ding, Y.~Yao, J.~Wang, Q.~Niu,
  \emph{Physical Review B}  \textbf{2010}, \emph{82}, 161414,
  \doi{10.1103/PhysRevB.82.161414}.

\bibitem{tse_quantum_2011}
W.-K. Tse, Z.~Qiao, Y.~Yao, A.~H. MacDonald, Q.~Niu, \emph{Physical Review B}
  \textbf{2011}, \emph{83}, 155447, \doi{10.1103/PhysRevB.83.155447}.

\bibitem{weeks_engineering_2011}
C.~Weeks, J.~Hu, J.~Alicea, M.~Franz, R.~Wu, \emph{Physical Review X}
  \textbf{2011}, \emph{1}, 021001, \doi{10.1103/PhysRevX.1.021001}.

\bibitem{hu_giant_2012}
J.~Hu, J.~Alicea, R.~Wu, M.~Franz, \emph{Physical Review Letters}
  \textbf{2012}, \emph{109}, 266801, \doi{10.1103/PhysRevLett.109.266801}.

\bibitem{zhang_electrically_2012}
H.~Zhang, C.~Lazo, S.~Bl{\"u}gel, S.~Heinze, Y.~Mokrousov, \emph{Physical
  Review Letters}  \textbf{2012}, \emph{108}, 056802,
  \doi{10.1103/PhysRevLett.108.056802}.

\bibitem{chandni_transport_2015}
U.~Chandni, E.~A. Henriksen, J.~P. Eisenstein, \emph{Physical Review B}
  \textbf{2015}, \emph{91}, 245402, \doi{10.1103/PhysRevB.91.245402}.

\bibitem{jia_transport_2015}
Z.~Jia, B.~Yan, J.~Niu, Q.~Han, R.~Zhu, D.~Yu, X.~Wu, \emph{Physical Review B}
  \textbf{2015}, \emph{91}, 085411, \doi{10.1103/PhysRevB.91.085411}.

\bibitem{efetov_controlling_2010}
D.~Efetov, P.~Kim, \emph{Physical Review Letters}  \textbf{2010}, \emph{105},
  256805, \doi{10.1103/PhysRevLett.105.256805}.

\bibitem{nakada_dft_2011}
K.~Nakada, A.~Ishii, in J.~R. Gong, editor, \emph{Graphene {{Simulation}}},
  1--20, {InTechOpen}, {Rijeka} \textbf{2011}.

\bibitem{manade_transition_2015}
M.~Manad{\'e}, F.~Vi{\~n}es, F.~Illas, \emph{Carbon}  \textbf{2015}, \emph{95},
  525, \doi{10.1016/j.carbon.2015.08.072}.

\bibitem{pi_electronic_2009}
K.~Pi, K.~M. McCreary, W.~Bao, W.~Han, Y.~F. Chiang, Y.~Li, S.-W. Tsai, C.~N.
  Lau, R.~K. Kawakami, \emph{Physical Review B}  \textbf{2009}, \emph{80},
  075406, \doi{10.1103/PhysRevB.80.075406}.

\bibitem{alemani_effect_2012}
M.~Alemani, A.~Barfuss, B.~Geng, {\c C}.~O. Girit, P.~Reisenauer, M.~F.
  Crommie, F.~Wang, A.~Zettl, F.~Hellman, \emph{Physical Review B}
  \textbf{2012}, \emph{86}, 075433, \doi{10.1103/PhysRevB.86.075433}.

\bibitem{katoch_impact_2012}
J.~Katoch, M.~Ishigami, \emph{Solid State Communications}  \textbf{2012},
  \emph{152}, 60, \doi{10.1016/j.ssc.2011.11.003}.

\bibitem{wang_neutral-current_2015}
Y.~Wang, X.~Cai, J.~{Reutt-Robey}, M.~S. Fuhrer, \emph{Physical Review B}
  \textbf{2015}, \emph{92}, 161411, \doi{10.1103/PhysRevB.92.161411}.

\bibitem{wang_electronic_2015}
Y.~Wang, S.~Xiao, X.~Cai, W.~Bao, J.~{Reutt-Robey}, M.~S. Fuhrer,
  \emph{Scientific Reports}  \textbf{2015}, \emph{5}, 15764,
  \doi{10.1038/srep15764}.

\bibitem{khademi_alkali_2016}
A.~Khademi, E.~Sajadi, P.~Dosanjh, D.~A. Bonn, J.~A. Folk, A.~St{\"o}hr,
  U.~Starke, S.~Forti, \emph{Physical Review B}  \textbf{2016}, \emph{94}, ISSN
  2469-9950, 2469-9969, \doi{10.1103/PhysRevB.94.201405}.

\bibitem{elias_electronic_2017}
J.~A. Elias, E.~A. Henriksen, \emph{Physical Review B}  \textbf{2017},
  \emph{95}, 075405, \doi{10.1103/PhysRevB.95.075405}.

\bibitem{huang_reliable_2015}
Y.~Huang, E.~Sutter, N.~N. Shi, J.~Zheng, T.~Yang, D.~Englund, H.-J. Gao,
  P.~Sutter, \emph{ACS Nano}  \textbf{2015}, \emph{9}, 10612, ISSN 1936-0851,
  1936-086X, \doi{10.1021/acsnano.5b04258}.

\bibitem{lin_graphene_2012}
Y.-C. Lin, C.-C. Lu, C.-H. Yeh, C.~Jin, K.~Suenaga, P.-W. Chiu, \emph{Nano
  Letters}  \textbf{2012}, \emph{12}, 414, \doi{10.1021/nl203733r}.

\bibitem{goossens_mechanical_2012}
A.~M. Goossens, V.~E. Calado, A.~Barreiro, K.~Watanabe, T.~Taniguchi, L.~M.~K.
  Vandersypen, \emph{Applied Physics Letters}  \textbf{2012}, \emph{100},
  073110, \doi{10.1063/1.3685504}.

\bibitem{lindvall_cleaning_2012}
N.~Lindvall, A.~Kalabukhov, A.~Yurgens, \emph{Journal of Applied Physics}
  \textbf{2012}, \emph{111}, 064904, \doi{10.1063/1.3695451}.

\bibitem{langmuir_vapor_1913}
I.~Langmuir, \emph{Physical Review}  \textbf{1913}, \emph{2}, 329,
  \doi{10.1103/PhysRev.2.329}.

\bibitem{noauthor_courtesy_nodate}
\emph{Courtesy {{SPI Supplies}}.}

\bibitem{adam_self-consistent_2007}
S.~Adam, E.~H. Hwang, V.~M. Galitski, S.~Das~Sarma, \emph{Proceedings of the
  National Academy of Sciences}  \textbf{2007}, \emph{104}, 18392.

\bibitem{hwang_single-particle_2008}
E.~H. Hwang, S.~Das~Sarma, \emph{Physical Review B}  \textbf{2008}, \emph{77},
  195412, \doi{10.1103/PhysRevB.77.195412}.

\bibitem{das_sarma_single-particle_1985}
S.~Das~Sarma, F.~Stern, \emph{Physical Review B}  \textbf{1985}, \emph{32},
  8442.

\bibitem{yan_correlated_2011}
J.~Yan, M.~S. Fuhrer, \emph{Physical Review Letters}  \textbf{2011},
  \emph{107}, 206601, \doi{10.1103/PhysRevLett.107.206601}.

\bibitem{barth_transport_2000}
J.~Barth, \emph{Surface Science Reports}  \textbf{2000}, \emph{40}, 75, ISSN
  01675729, \doi{10.1016/S0167-5729(00)00002-9}.

\bibitem{honig_vapor_1969}
R.~E. Honig, D.~A. Kramer, \emph{RCA Review}  \textbf{1969}, \emph{30}, 285.

\bibitem{katsnelson_scattering_2009}
M.~I. Katsnelson, F.~Guinea, A.~K. Geim, \emph{Physical Review B}
  \textbf{2009}, \emph{79}, 195426, \doi{10.1103/PhysRevB.79.195426}.

\bibitem{mccreary_effect_2010}
K.~M. McCreary, K.~Pi, A.~G. Swartz, W.~Han, W.~Bao, C.~N. Lau, F.~Guinea,
  M.~I. Katsnelson, R.~K. Kawakami, \emph{Physical Review B}  \textbf{2010},
  \emph{81}, 115453, \doi{10.1103/PhysRevB.81.115453}.

\bibitem{cresti_multiple_2014}
A.~Cresti, D.~Van~Tuan, D.~Soriano, A.~W. Cummings, S.~Roche, \emph{Physical
  Review Letters}  \textbf{2014}, \emph{113}, 246603, ISSN 0031-9007,
  1079-7114, \doi{10.1103/PhysRevLett.113.246603}.

\bibitem{schedin_detection_2007}
F.~Schedin, A.~Geim, S.~Morozov, E.~Hill, P.~Blake, M.~I. Katsnelson, K.~S.
  Novoselov, \emph{Nature Materials}  \textbf{2007}, \emph{6}, 652.

\end{thebibliography}

\end{document}